# Time-Stretch Probing of Ultrafast Soliton Molecule Dynamics

Shuqian Sun[1,2], Zhixing Lin[1,3], Wei Li[1,2], Ninghua Zhu[1,2],* & Ming Li[1,2],*

[1]State Key Laboratory of Integrated Optoelectronics, Institute of Semiconductors, Chinese Academy of Sciences, Beijing 100083, China
[2]School of Electronic, Electrical and Communication Engineering, University of Chinese Academy of Science, Beijing 100049, China
[3]College of Materials Science and Opto-Electronic Technology, University of Chinese Academy of Science, Beijing 100049, China
*Corresponding to: ml@semi.ac.cn, nhzhu@semi.ac.cn

**Abstract:**
**Soliton molecules are the manifestation of attractive and repulsive interaction between optical pulses mediated by a nonlinear medium. However, the formation and breakup of soliton molecules are difficult to observe due to the transient nature of the process. Using the time stretch technique, we have been able to track the real-time evolution of the bound state formation in a femtosecond fiber laser and unveil different evolution paths towards a stable bound state. A non-stationary Q-switched mode-locking regime consisting of various transient solitons is observed in this transition period. We also observe additional dynamics including soliton molecule vibrations, collision and decay. Our findings uncover a diverse set of soliton dynamics in an optical nonlinear system and provide valuable data for further theoretical studies.**

## Introduction:

Mode-locked laser is one of the most attractive research area in optical system, not only for its important applications in nanomachining, optical communications and medical operation[1,2], but also for its unique value in scientific research. As a typical nonlinear system, mode-locked laser is a great research platform for optical solitons[3,4], a localized optical field balanced by dispersion and nonlinear effects. Specially in a dissipative system, solitons can display a much richer range of behaviors account for another balance between energy gain and loss[4,5]. Apart from single soliton operation, soliton bound states consisting of multi-pulse is theoretically described[6-8] and experimentally observed[9-11]. Based on the complex Ginzburg-Landau equation (CGLE), more complicated soliton dynamics are predicated[3,12-15], including pulsating solitons, chaotic solitons, soliton explosion, bound state oscillations and collisions. However, the evolution details of many unstable states have not been revealed in real time limited by the measurement method.

Recently, the development of Time-Stretch Dispersive Fourier Transform (TS-DFT) technique makes us possible to trace the real-time spectral evolution of a transient event[16-18]. Using a dispersion medium, spectral information is transferred to the time domain and is slowed down so it can be captured in real time with an electronic analog-to-digital converter[19]. Thus the whole system can serve as a real-time optical spectrum analyzer (OSA). The simple but powerful method has been used to resolve the establishment of mode locking in a Kerr-lens mode-locking Ti:sapphire laser[20], as well as the soliton bound state dynamics inside it[4]. This method is also used in the observation of transient solitons, soliton explosion and other soliton dynamics with complex interactions[21-25].

Using the TS-DFT technique, we observed the real-time evolution of the soliton molecule formation in a passively mode-locked fiber laser and found fascinating and previously unobserved dynamics. Since the upper-state lifetime of Er-doped fiber (EDF) is much longer than Ti: sapphire[26], the soliton dynamics in a fiber laser shows quite different behaviors compared to the Ti: sapphire mode-locked laser. Rather than direct evolution from a single pulse operation[4], the formation of soliton bound state in the fiber laser undergoes a non-stationary Q-switched mode-locking regime, where a variety of transient solitons emerge and collapse. For the first time, we resolve the bound state formation from a Q-switched burst and find several different evolution paths. Bound state oscillations and collisions before dissipation are also recorded and analyzed. Although the transient Q-switched regime prior to a stable mode locking has been observed by many researchers [21,27-32], it has never been found between the single-pulse operation and the soliton molecule regime. These transient solitons show more complicated dynamics compared to the results in [21].

**Results**

The laser used in our experiment is a commercial PriTel femtosecond fiber laser with 50 MHz repetition rate. The block diagram of the laser is shown in Fig. 1a. Although we usually use the laser at single-pulse operation, this system can also support double pulse regime. The spectrum in this regime has interferometric fringe under a Gaussian envelope (see Supplementary Materials Fig. S1). Output pulses of the laser are stretched by a length of dispersion compensation fiber (DCF) and then the spectral information is mapped into time domain (see Fig. 1b). Thus we can monitor the real-time spectral evolution using an oscilloscope (see Method).

To prepare a soliton bound state, we increase the pump current beyond a critical level. The critical value may vary in different measurement but is always beyond 460 mA. The output spectrum will evolve gradually with the increasing pump current (see Supplementary Materials Fig. S1). However, when the pump current reaches the threshold value, the laser will go through a transient Q-switched mode-locking regime before entering the stable soliton bound state. This transition regime usually sustains several hundred microseconds. Sometimes it can be longer than 1 s (see Supplementary Materials Fig. S1). Q-switched mode locking has been experimentally reported in many mode-locked laser systems [21,27-32], but has never been observed between the single-pulse regime and the soliton molecule regime. This phenomenon is related to the saturable absorption and the relaxation oscillation inside the resonant cavity[29,32]. It is worth noting that the Q-switched mode-locking regime can be divided into many groups, each of which consists of a train of bursts (see Supplementary Materials Fig. S1). The gap between groups is actually an unstable single-pulse state.

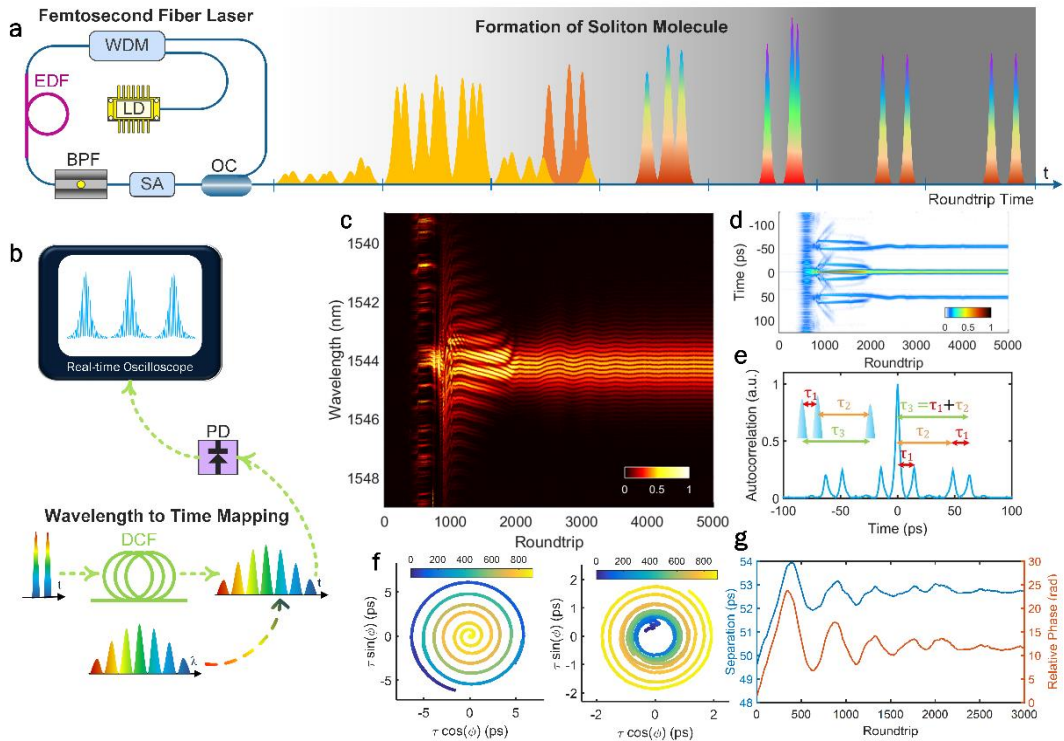

**Fig. 1. The formation of soliton molecule in a femtosecond fiber laser.** (**a**) A stable soliton molecule evolves from a Q-switched mode-locking burst. Left is the block diagram of the femtosecond fiber laser used in our experiment. The pump source is a laser diode (LD) at 980 nm. A spool of Er-doped fiber (EDF) is used as the gain medium. WDM: wavelength division multiplexer, BPF: band pass filter, SA: saturable absorber, OC: optical coupler. (**b**) The experimental setup of TS-DFT technique. The output pulse is time-stretched in a 10-km dispersion compensation fiber (DCF) and its spectrum is mapped in temporal domain. The stretched pulse is obtained by a high-speed photodetector and monitored in the real-time oscilloscope. (**c**) Experimental real-time spectral evolution of the formation of a soliton molecule from the last Q-switched mode-locking burst. (**d**) The Fourier transform of the measured real-time spectrum, representing the autocorrelation of its temporal waveform. (**e**) The autocorrelation of the 1500th roundtrip. The inset shows the position relation of the three solitons. (**f**) Left: The interaction plane of pulse1 and pulse2 over roundtrips from 1000 to 1900. The radius (the center is 10 ps) and angle represent the pulse separation and relative phase, respectively. The clockwise trajectory indicates a monotonically decreasing relative phase and a reduced separation. Right: The interaction plane of pulse2 and pulse3 over roundtrips from 1000 to 1900. The trajectory is counterclockwise, indicating an increasing relative phase with an expanded peak separation (the center is 47 ps). (**g**) The separation and relative phase as a function of roundtrips over roundtrips from 2000 to 5000, illustrating a decaying fluctuation before entering a stable bound state.

## Formation, vibration and decaying of stable soliton molecules

The stable soliton molecules only emerge after the transient Q-switched mode-locking regime. Fig. 1c traces the formation of a stable bound state from the last Q-switched burst. As shown, a spectrum with superimposed fringe periodicities emerges within the Q-switched burst, which implies a triplet soliton arises in this burst. This can be verified by the temporal autocorrelation shown in Fig. 1d. A transient triplet soliton is created in the Q-switched burst before evolving to a stable doublet soliton. Fig. 1e shows the autocorrelation of the 1500th roundtrip, indicating the ubiety of the three pulses. During the following 1000 roundtrip, the spectrum range shrinks from ~8 nm to ~2 nm and two pulses of the triplet soliton attract each other gradually. At the end of the Q-switched burst, the triplet state decays into a doublet soliton at ~50 ps separation eventually.

The interaction plane is usually used to describe the behavior of multi-soliton, exhibiting the pulse separation and the relative phase in a polar diagram[6]. Pulse separations can be obtained from the Fourier transform of recorded real-time spectra, while the relative phase between two pulses is calculated from the shift of fringe patterns[4]. Fig. 1f (left) illustrates the interaction plane of pulse1 and pulse2 from 1000 to 1900 roundtrips. The clockwise trajectory in the interaction plane illustrates a decreasing relative phase with shrinking peak separation. The evolution of relative phase is related to the intensity difference of the two pulses[4]. Because of the Kerr effect inside the fiber, pulses with different intensities have different phase velocities, yielding a variation of relative phases. Fig. 1f (right) exhibits the interaction plane of pulse2 and pulse3 (the two pulses left in the stable bound state) from 1000 to 1900 roundtrips. Contrary to the interaction between pulse1 and pulse2, Fig. 1f (right) illustrates a counterclockwise trajectory with an increasing separation. Since pulse 2 is in the middle of the triplet soliton, the opposite variation of relative phases shown in Fig. 1f indicates that the intensity of pulse2 is the highest among the three pulses. After a monotonically increasing, the relative phase and binding separation undergo a fluctuation before reaching a locked relative phase and a constant separation of ~53 ps (see Fig. 1g). Based on the theory mentioned above, we expect that the intensity difference of the two solitons go through a similar variation. And in the stable regime, the intensities of two pulses should be the same. The binding separation and phase difference of the bound state is determined by the energy and momentum inside the system, which has been described theoretically [6].

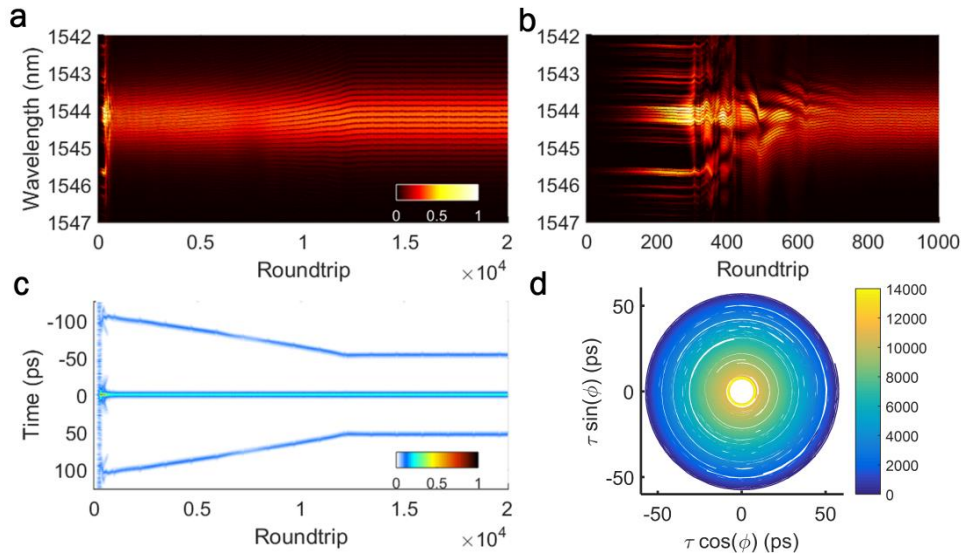

**Fig. 2. Another path of soliton molecule formation.** The real-time spectral evolution (**a**) and the corresponding temporal autocorrelation (**c**) of another bound state establishment event. Two solitons evolve from the Q-switched mode-locking burst and attract each other in the following roundtrip until a stable bound state is achieved. (**b**) The expanded view illustrates the very beginning of this formation event. An interferogram emerges from a short-period noise-like background pulses. (**d**) The corresponding interaction plane over roundtrips from 1000 to 15000, illustrating a clockwise trajectory with a reduced separation from ~100 ps to ~50 ps (the center is 45 ps).

Apart from the soliton molecule formation recorded above, we observed a different formation event shown in Fig. 2a. Here a Q-switched burst produce a pulse pair directly, rather than a triplet soliton. However, the pulse separation of this doublet, about 100 ps, is much larger than a stable bound state. In the following 12000 roundtrips, the two solitons attract each other gradually until a stable bound state with ~50 ps separation is achieved (see Fig. 2c). Fig. 2d exhibits the corresponding interaction plane over roundtrips from 1000 to 15000. Accompanied by a reduced binding separation, the relative phase of the two pulses decreases gradually, giving a clockwise trajectory. The establishment prior to the stable bound state covers over 12000 roundtrips, much longer than the bound state formation (less than 5000 roundtrips) in Fig. 1c.

Besides the two formation processes shown in Fig. 1 and 2, we recorded another qualitatively different case, which is shown in Fig. 3. In this formation event, a Q-switched mode-locking burst transfers into a momentary single pulse state firstly, as shown in Fig. 3a & b. In the following evolution, an interferogram arises gradually in the Gaussian-like spectrum, implying the emergence of the second pulse. Undergoing attraction and interaction, the soliton doublet settles at a separation around 50 ps (see Fig. 3d). Noting the complete establishment process of a stable bound state is usually more than 10 milliseconds in this case, which is difficult to record it in a single-shot measurement (the maximum record length at 25 GS/s sampling rate is 10 milliseconds). Therefor the real-time spectra shown in Fig. 3a&c are measured independently. It is worth noting that the corresponding autocorrelation shown in Fig. 3d illustrates a turn-back track, meaning that the second soliton emerges just next to the original pulse and then depart from each other before merging at 150 ps separation. However, the maximum separation (150 ps) is actually decided by the spectral resolution (limited by the sampling rate of the oscilloscope). The spectrum prior to the 38000[th] roundtrip is under-sampled, which means the second pulse should have emerged at 300 ps separation and draw near to each other continuously in the following 58,000 roundtrips.

Actually, the soliton molecule shown in Fig. 3 is a dynamic bound state, where binding separation and relative phase vibrate rapidly over time[14,15]. After about 15,000 roundtrips, this dynamic soliton molecule merges into one soliton and disappears eventually. The expanded view of soliton vibration is shown in Fig. 3e, in which violent oscillations and moderate vibrations alternate. The extracted relative phase and pulse separation are shown in Fig. 3f, where the relative phase vibrates over $2\pi$ and the binding separation oscillates around 52 ps. The soliton oscillations also show some periodicity, with repeat period about 500 roundtrips. Fig. 3g shows the expanded view of bound state mergence, with corresponding configuration space shown in Fig. 3h. After soliton oscillations, the two solitons undergo attraction over 5000 roundtrips with an increasing relative phase, colliding and decaying eventually.

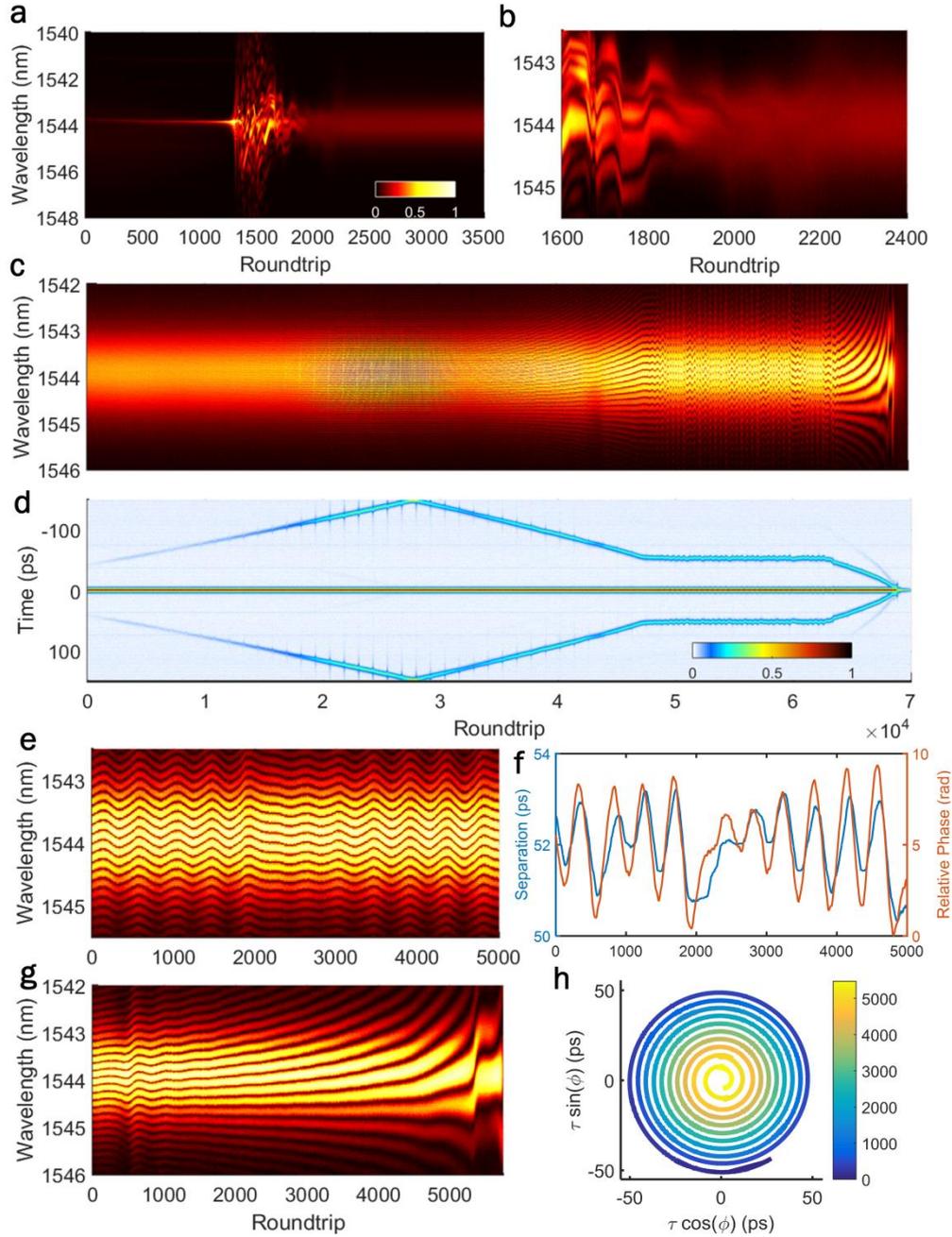

**Fig. 3. The formation, vibration and decaying of a soliton molecule.** (**a**) The spectral evolution from the last Q-switched mode-locking burst to a single pulse. (**b**) The expanded view of (**a**). (**c**) The real-time spectrum and (**d**) corresponding temporal autocorrelation of the formation of soliton molecules, as well as bound states vibrations and decaying. The expanded view (**e**) shows the details of soliton molecule vibration, which shows some periodicity with repeat period ~ 500 roundtrips. (**f**) The separation and relative phase as a function of roundtrips. (**g**) The real-time spectrum of bound state decaying in details. As shown the two solitons attract each other over 5000 roundtrips and decay eventually. (**h**) The corresponding interaction plane of soliton molecule decaying (the center is 0 ps).

## Unstable Soliton Molecules

From the recorded real-time spectra of Q-switched mode-locking burst, we found various unstable soliton molecules, some of which are shown in Fig. 4a-c. Although the transient soliton under a Q-switched envelope has been analyzed[21], we observe a richer range of soliton dynamics between single pulse regime and the bound state. We begin in Fig. 4a by showing the first burst of one group of Q-switched mode-locking bursts. This kind of bursts always start from a single wavelength and extend suddenly after hundreds of roundtrips. At the same time, irregular fringe patterns arise accompanied by power fluctuations. The Fourier transform of this spectrum shows that some disordered multi-pulses are created in this process. We record a large number of Q-switched bursts in our experiment but regular patterns are hardly found in this kind of bursts. Eventually, the Q-switched burst fades out gradually with spectrum shrinking.

Different from the first one, other bursts emerge after a noise-like pulse fluctuation, which is similar to the background fluctuation before mode-locking build-up mentioned in Ref.[20]. Regular fringe patterns often emerge in these Q-switched bursts. As shown in Fig. 4b, a clear spectral fringe can be observed after a noise-like period. Its temporal autocorrelation shows that a soliton pair separated by 15 ps is created in this transient burst. It is worth noting that the soliton separation remains almost unchanged in the following 900 roundtrips before decaying. A soliton consisting three of four pulses can also be observed in other bursts, showing a diverse set of dynamics (see Fig. 4c).

Next we discuss a quite different Q-switched burst shown in Fig. 4d. Here a doublet soliton with ~10 ps separation emerges firstly. In the following 2000 roundtrips, the two pulses attract each other gradually until reaching a metastable bound state separated by 5 ps. But this metastable soliton pair only sustains for about 1000 roundtrips before merging into one pulse and decaying eventually. We even observed a transient burst consisting of two sequential doublet solitons with extremely long lifetime (about 8000 roundtrips), as shown in Fig. 4e. The second soliton pair emerges just after the colliding of the first one, with larger separation and longer lifetime. Commonly, this kind of transient soliton is the last one in a bursts group with a lifetime several times longer than other Q-switched bursts.

The emergence of Q-switched bursts shows some stochasticity[33]. Based on the theory described in [6,34], the parameter regions supporting soliton bound state is much narrower than those supporting a stable single soliton. So the transient Q-switched mode-locking regime can be considered as a searching period for suitable parameters. Numbers of transient solitons collapse during Q-switching regime until a candidate with suitable separation and phase arises. Then this soliton will evolve into a stable bound state.

## Discussion

By means of the TS-DFT technique, we resolve the formation dynamics and interactions of soliton molecules inside a femtosecond fiber laser. Different from the Ti: sapphire mode-locked laser, the upper-state lifetime of EDF is much longer so that the soliton dynamics in a fiber laser shows quite different behaviors. A rich and fascinating variety of transient solitons under Q-switched envelopes prior to the stable bound state were tracked and analyzed. Specially, we found that the stable soliton molecule can evolve from Q-switched bursts in several different paths, which has never been observed before. The bound state vibration and decaying were also resolved in real time. These findings reveal a panorama of the soliton molecule evolution, significantly improving our understanding of complex dynamics in nonlinear systems.

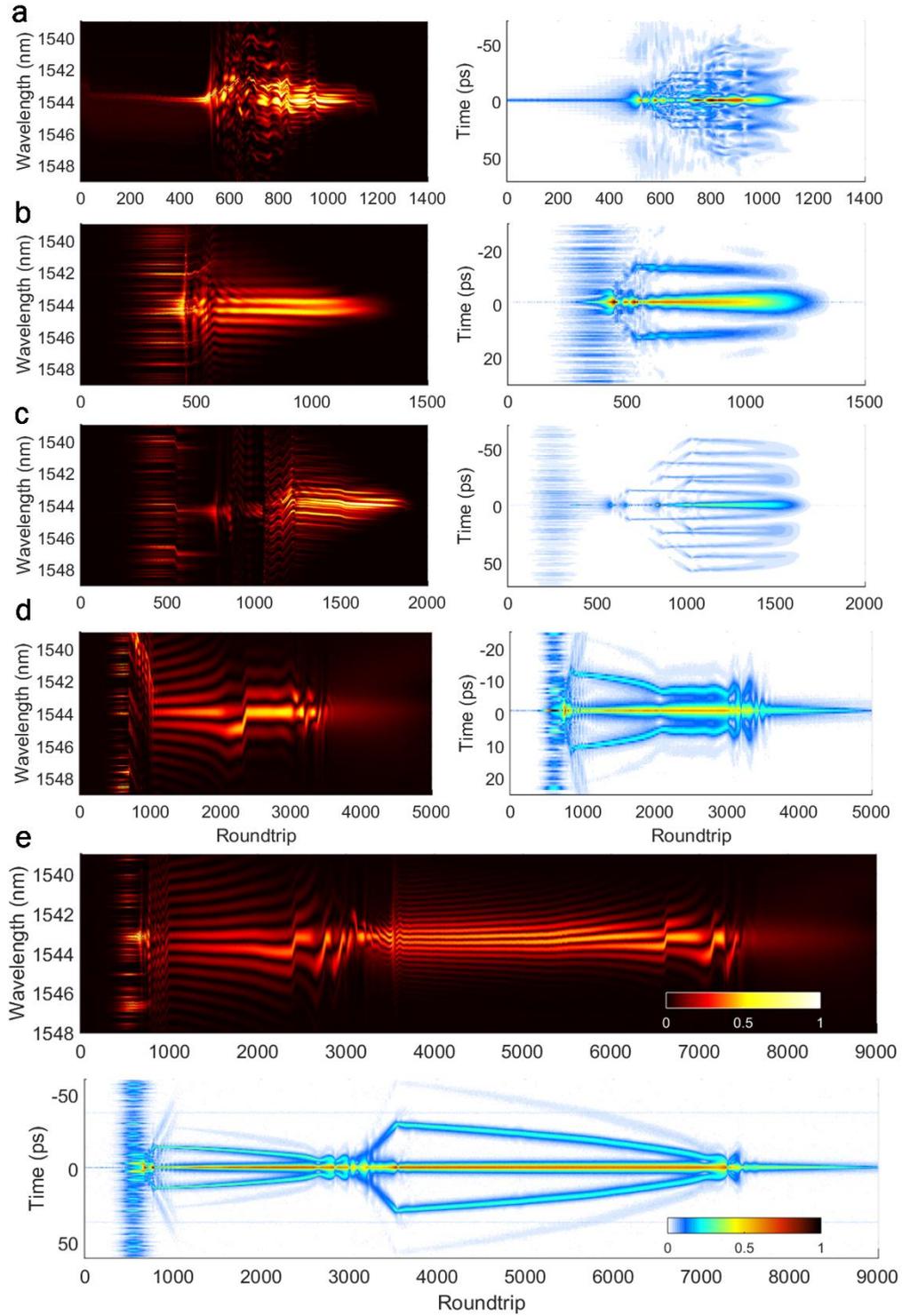

**Fig. 4. The real-time evolution of several typical Q-switched mode-locking bursts.** (**a**) The first one in a group of Q-switched mode-locking burst. It emerges from a narrow spectrum and then broaden suddenly after several hundred roundtrips with complex fringe patterns. (**b**) and (**c**) illustrate the bursts inside a Q-switched burst group. These transient solitons always arise after a short-term noise-like period and evolve into a regular multi-soliton in the following roundtrips. The lifetime of these transient solitons are usually less than 1000 roundtrips. (**d**) and (**e**) reveal the soliton dynamics in the end of a Q-switched mode-locking group, showing an obvious attraction and collision. The lifetime of these transient solitons are often several times longer than others.

## Methods

The laser used in our experiment is a commercial PriTel femtosecond fiber laser with 50 MHz repetition rate. The pulse width at stable single-pulse operation is about 120 fs. To record the real-time spectral evolution of soliton dynamics, we use a 10-km DCF with dispersion value of 1486 ps/nm to stretch the ultra-fast pulse. The stretched optical pulse is then fed into a high-speed photodector (PD) with 18 GHz bandwidth and monitored in a real-time oscilloscope (Tektronix, MSO 71604C Mixed Signal Oscilloscope). The sampling rate of the oscilloscope is set to be 25 GS/s.

**Acknowledgement:**

This research is supported by National Natural Science Foundation of China under 61377002, 61321063 and 61090391. Ming Li was also supported by Thousand Young Talent program. Furthermore, the authors would like to thank Prof. Bahram Jalali for his great advice in paper writing.

**Author Contributions**

Shuqian Sun operates the experiment and writes the paper. Zhixing Lin participates in the experimental work. Wei Li and Ming Li review the manuscript. Ming Li and Ninghua Zhu provide the overall supervision.

**Competing financial interests**

The authors declare no competing financial interests.

# Time-Stretch Probing of Ultrafast Soliton Molecule Dynamics：

## Supplementary Materials

Shuqian Sun[1,2], Zhixing Lin[1,3], Wei Li[1,2], Ninghua Zhu[1,2],* & Ming Li[1,2],*

[1]State Key Laboratory of Integrated Optoelectronics, Institute of Semiconductors, Chinese Academy of Sciences, Beijing 100083, China
[2]School of Electronic, Electrical and Communication Engineering, University of Chinese Academy of Science, Beijing 100049, China
[3]College of Materials Science and Opto-Electronic Technology, University of Chinese Academy of Science, Beijing 100049, China
*Corresponding to: ml@semi.ac.cn, nhzhu@semi.ac.cn

### Spectra of Single Pulse Operation and Soliton Molecule Regime

The single-pulse operation has a Gaussian-like spectrum, which is shown in Fig. S1a (blue). When the pump current is increased beyond a critical value, the laser will enter a soliton molecule regime, in which the output becomes a stable pulse pair with a separation of ~50 ps. The spectrum of soliton molecules at 465 mA is an interferogram under a Gaussian envelope, where the fringe period reflects the soliton separation (shown in green in Fig. S1a). When we increase the pump current from 370 mA, the spectrum evolves gradually before the critical level, enabling us to record the spectral evolution using an OSA (see Fig. S1b). Each trace in Fig. S1b illustrates the spectrum at different pump current. Upon increment of pump level, the spectral width is extended gradually and the spectrum sinks bit by bit at ~1544 nm because of the nonlinearity in the femtosecond fiber laser. However, a further increment of the pump current will stimulate a narrow peak at around 1544 nm, which is likely a continuous wave component[1].

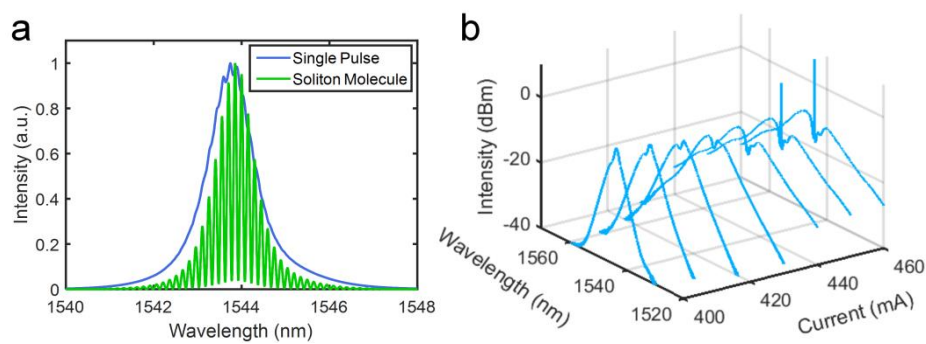

**Fig. S1.** (**a**) Measured spectra of single pulse regime at 370 mA (blue) and soliton molecule regime at 465 mA (green). (**b**) The optical spectral evolution recorded by OSA before entering soliton molecule regime.

### Q-switched Mode-locking Burst

When the current is increased beyond the critical level, the laser will go through a fleeting Q-switch mode-locking regime and then enter a stable bound state. The repetition period of the Q-switched bursts is about 60 microseconds, with a temporal width ranging from several microseconds to tens of microseconds. This transition regime usually sustains several hundred microseconds. Sometimes it can also be longer than 1 s. As shown in Fig. S2a, the single pulse operation and the soliton molecule state are divided by a ~1.2 s Q-switched mode-locking regime. It is worth noting that the Q-switched mode-locking regime can be divided into many such groups, each of which consists of several transient bursts. The gap between groups is usually hundreds of microsecond and the optical power inside it is not zero. Fig. S2b shows one group of Q-switched mode-locking bursts. The expanded view of Fig. S2b (see Fig. S2c and Fig. S2d) shows a train of time-stretched unstable mode-locked pulses at 20 ns period, corresponding to the repeat rate of 50 MHz. The temporal waveform represents its spectral information because of dispersive Fourier transform employed in the experiment.

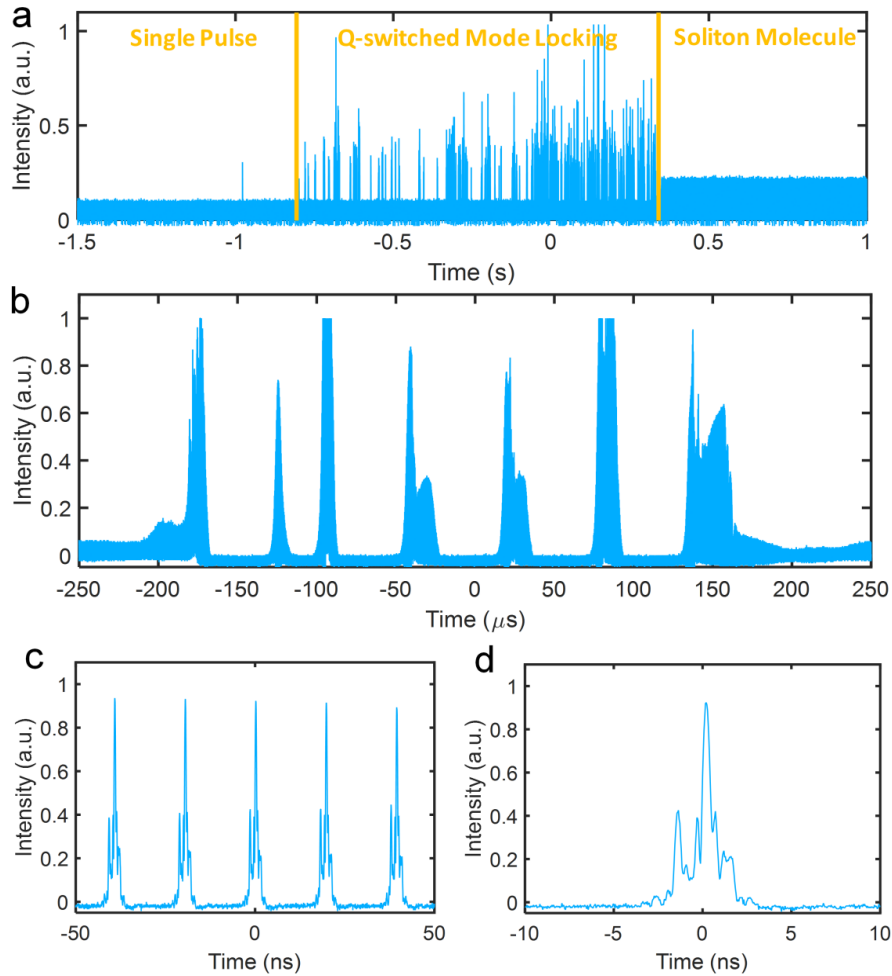

**Fig. S2.** (**a**) The recorded transient regime from single pulse regime to soliton bound state over a time-window of 2.5 s, in which a ~1.2 s Q-switched mode-locking regime is illustrated. (**b**) One group of Q-switched mode-locking bursts repeating at about 60 microseconds period. The temporal width of these bursts are not the same, ranging from several microseconds to tens of microseconds. (**c**) The zoom-in view of (**b**) shows unstable mode-locked pulses at 20 ns period (corresponding to 50 MHz repeat rate of the laser) under each Q-switched mode-locking burst. (**d**) The temporal waveform represents its spectral information because of dispersive Fourier transform employed in the experiment.